\documentclass[twocolumn,letterpaper,aps,prl,superscriptaddress,showpacs,floatfix]{revtex4}

\usepackage{graphicx}	
\usepackage{color}
\usepackage{bm} 

\usepackage{xspace}	

\usepackage{grffile} 

\newcommand{\dphi}{\mbox{$\Delta\phi$}\xspace}

\newcommand{\pt}{\mbox{$p_T$}\xspace}

\newcommand{\raa}{\mbox{$R_{\rm AA}$}\xspace}

\newcommand{\pvec}       {\mathbf{p}}
\newcommand{\xivec}      {\bm{\xi}}
\newcommand{\ccbar} {c\bar{c}}
\newcommand{\bbbar} {b\bar{b}}
\newcommand{\qqbar} {q\bar{q}}
\begin{document}


\title{Tests of the Quark-Gluon Plasma Coupling Strength at Early Times with Heavy Quarks}

%

\author{A. M. Adare}
\affiliation{University of Colorado at Boulder}
\author{M. P. McCumber}
\affiliation{Los Alamos National Laboratory}
\author{J. L. Nagle}
\affiliation{University of Colorado at Boulder}
\author{P. Romatschke}
\affiliation{University of Colorado at Boulder}

\date{\today}

\begin{abstract}
  The redistribution in momentum space of heavy quarks via their
  interactions in the quark-gluon plasma is an excellent probe of the
  heavy quark coupling strength to the medium. We utilize a Monte
  Carlo Langevin calculation for tracking heavy quark - antiquark
  pairs embedded in a viscous hydrodynamic space-time evolution. We
  find that the nuclear modification factor ($R_{AA}$) for charm
  quarks is relatively insensitive to the coupling to the quark-gluon
  plasma at early times where the highest temperatures are
  acheived. In contrast the azimuthal angular correlation of charm and
  anticharm quarks is extremely sensitive to the early time evolution.
  For beauty quarks the situation is reversed in terms of sensitivity.
  Future measurements of heavy quarks have the potential to map
  out the temperature dependence of the shear viscosity to entropy
  density ratio ($\eta/s$).
\end{abstract}


\maketitle

High energy heavy ion collisions at the Relativisitic Heavy Ion
Collider (RHIC) and the Large Hadron Collider (LHC) produce nuclear
matter at sufficiently high temperatures to create droplets of the
quark-gluon plasma (QGP). Even at the highest temperatures achieved,
thermal production of heavy quark-antiquark pairs is suppressed and
the $c\overline{c}$ and $b\overline{b}$ pairs are produced primarily
at the earliest times in large momentum transfer reactions between
incoming partons within the incident nuclei. Due to flavor
conservation of the strong interaction, the heavy quarks emerge from
the QGP within a charm or beauty hadron. Heavy quarks therefore act as
``tracers'' that record the evolution of the QGP through thermalization,
hydrodynamic expansion, and hadronization, even if the QGP itself has
no long-lived quasiparticles~\cite{nsac2012}.

It was proposed in Ref.~\cite{Dokshitzer:2001zm} that the interactions
of heavy quarks at modest transverse momenta ($p_{T} < M_{Q}$) would
have a weaker effective coupling to the medium by a ``dead cone''
effect that reduces the phase space for radiative energy loss.
However, initial experimental results were consistent with charm
quarks following the flow of the underlying quark-gluon
plasma~\cite{Batsouli:2002qf}.  Subsequently, the degree of
thermalization was studied within a Langevin approach by Moore and
Teaney~\cite{Moore:2004tg}. Reasonable agreement with the suppression
and elliptic flow of heavy quark mesons measured via semi-leptonic
decay electrons is achieved with a diffusion rate requiring the shear
viscosity over entropy density ($\eta/s$) to lie within a factor of
two of the conjectured $1/4\pi$ limit~\cite{Adare:2006nq}. Numerous
works have employed similar Langevin calculations with different
assumptions about the underlying quark-gluon plasma space-time
evolution
~\cite{He:2013zua,Cao2013653c,Cao:2012as,Xu:2013uza,Akamatsu:2008ge,Zhu:2007ne}.

\begin{figure}[bt]
  \includegraphics[width=\linewidth]{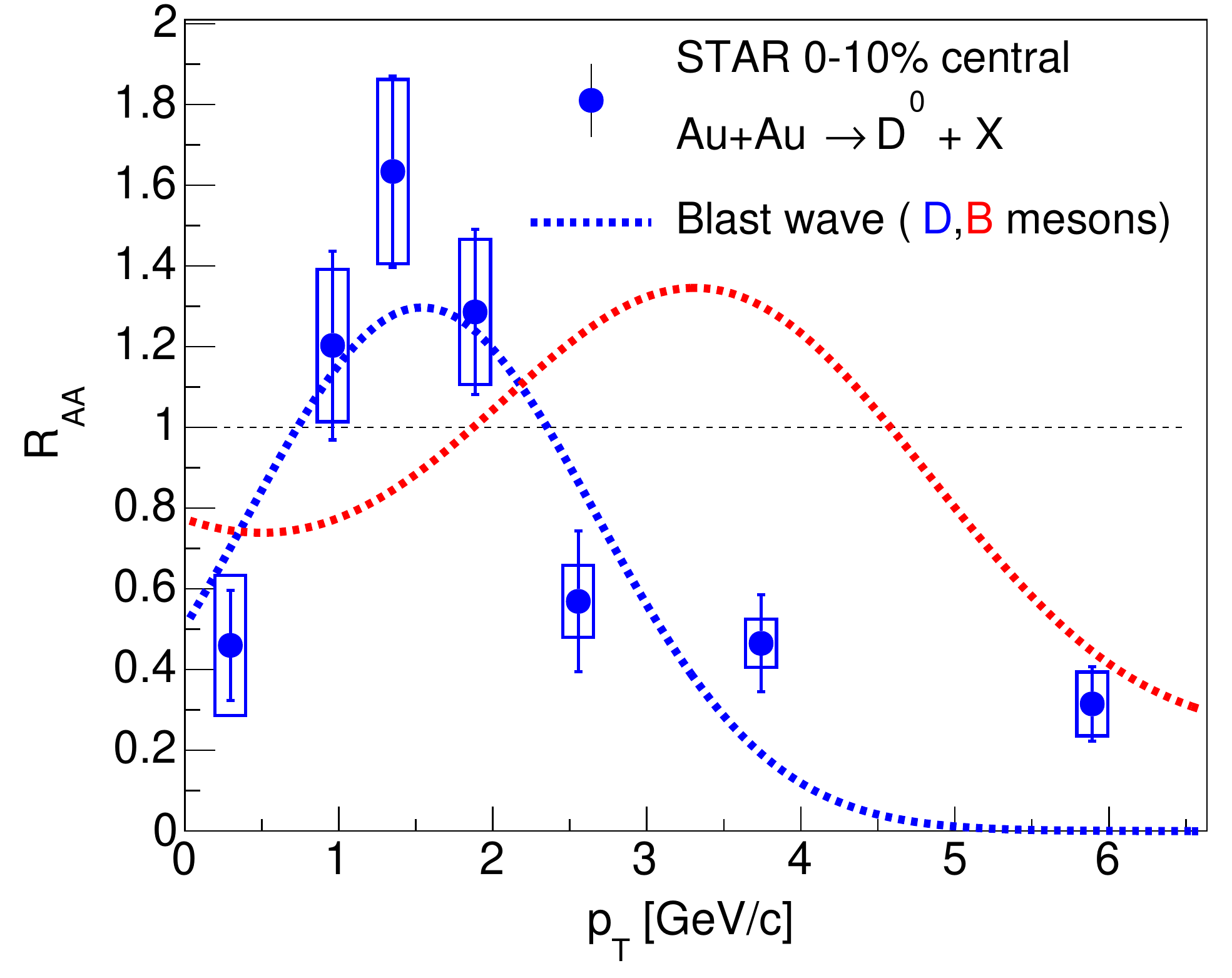}
  \caption{\label{fig:blast} $\raa$ of $D^0$ mesons in 0-10\% central
    $Au+Au$ at $\sqrt{s_{_{{NN}}}} = 200$ GeV compared with blast-wave
    calculations for D and B mesons and a PYTHIA $p+p$ baseline
    reference.}
\end{figure}

A preliminary measurement of the $D$ meson $R_{AA}$ in 0-10\% central
$Au+Au$ collisions at $\sqrt{s_{_{{NN}}}} = 200$ GeV is shown in
Figure~\ref{fig:blast}~\cite{Tlusty:2012ix}. The data are in close
agreement with a blast wave prediction from
Ref.~\cite{Batsouli:2002qf} up to $p_{T} \approx 3.5$ GeV/$c$. The
calculation utilizes PYTHIA for the $p+p$ baseline, and for Au+Au a
linear boost profile blast wave model constrained by $\pi$, $K$, and
$p$ transverse momentum distributions.  In the blast wave model, an
outward push from radial flow leads to a trend of suppression for
$p_{T} < 1$ GeV/$c$, followed by an enhancement for $p_{T} \approx
1-2.2$ GeV/$c$. Since the radial flow boost available within the model
is limited, suppression occurs for $p_{T} > 2.2$ GeV/c. At higher
momenta, the heavy quarks increasingly deviate from thermal
equilibrium, and the blast wave model and the data are expected to
diverge. 
In this paper, we aim to understand the full time evolution of
the charm and beauty quark distributions in space and momentum, and to
test whether the blast wave final-state parametrization is
reproducible.

We have implemented a Monte Carlo Langevin calculation to trace the
diffusion and drag of individual heavy quarks. We have tested the
numerical algorithm against the control thermalization tests in
Ref.~\cite{Cao:2012as} and obtain identical results. The transverse
momentum distribution of the initial heavy quarks are selected from
the following equation:
\begin{equation}
  \frac{1}{\pt}\frac{dN}{d\pt} \propto \frac{1}{(\pt^2 +
    \Lambda^2)^{\alpha}}
\end{equation}
where $\alpha = 3.9 \, (4.9)$ and $\Lambda = 2.1 \, (7.5)$ for charm
(beauty) quarks, following Ref.~\cite{Cao:2012jt}.  We then generate
initial conditions by averaging over central $Au+Au$ collisions at
$\sqrt{s_{NN}} = 200$ GeV with a Monte Carlo Glauber
code~\cite{Alver:2008aq}, where each event is rotated into the axis of
the participant plane. The event averaging ensures a smooth spatial
configuration for numerical stability in the subsequent hydrodynamic
evolution. The initial heavy quark-antiquark pair positions are
sampled from this smooth distribution of binary collisions.

An initial transverse momentum $k_T$ sampled randomly from
$\mathcal{N}(k_T \,|\, \mu = 0, \sigma^2 = 1.0 \,\, \mathrm{GeV}/c)$
is added to the $\ccbar$ and $\bbbar$ pairs at their point of
production, where $\mathcal{N}$ is the normal distribution. We note
that the effect of varying $k_T$ has been studied in
e.g. Ref.~\cite{Xu:2013uza}, and in the end such parameters must be
constrained from $p+p$ and $p(d)+A$ experimental data.

We then run the viscous hydrodynamic code from Luzum and
Romatschke~\cite{PhysRevC.78.034915,PhysRevC.79.039903} to generate
the space-time distribution of temperature $T$, energy density, and
fluid velocities. The original code has been modified for new input
and output formats. We then run individual heavy quark-antiquark pairs
in time steps of 0.025 fm/c through the space-time background
distribution, updating the 3-momentum information at each step
according to the Langevin equation
\begin{equation}
  \frac{d \pvec(t)}{dt} = - \eta_D(p)\,\pvec(t) +  \xivec(t).
\end{equation}
As a consequence of the fluctuation-dissipation theorem, all of the
essential physical effects (scattering, viscous drag, and hydrodynamic
boosts) are controlled by a single diffusion parameter $D$ at the
local temperature $T$ of the thermal background, under the assumption
of small individual energy transfers. The viscous drag force
$\eta_D(p) = T/(ED)$ describes large-scale average motion of a
particle with energy $E \approx M$, while $\xi^{i}$ describes
fluctuations in coordinate $i$ about the average motion as follows:
\begin{equation}
  \langle \xi^i(t) \, \xi^j(t') \rangle =  %
  \frac{2T^2}{D}\delta^{ij} \delta(t-t'), \qquad %
  \langle \xi^i(t) \rangle = 0.
\end{equation}
This is implemented in the Langevin calculation by applying a momentum
deflection $\Delta p$ sampled at random from $\mathcal{N}(\Delta p
\,|\, \mu = 0, \sigma^2 = 2T^2 / D \Delta t)$ at each time step
$\Delta t$.  We tested that increasing $\Delta t$ by a factor of 10
does not change the results.
\begin{figure}[htb]
  \includegraphics[width=\linewidth]{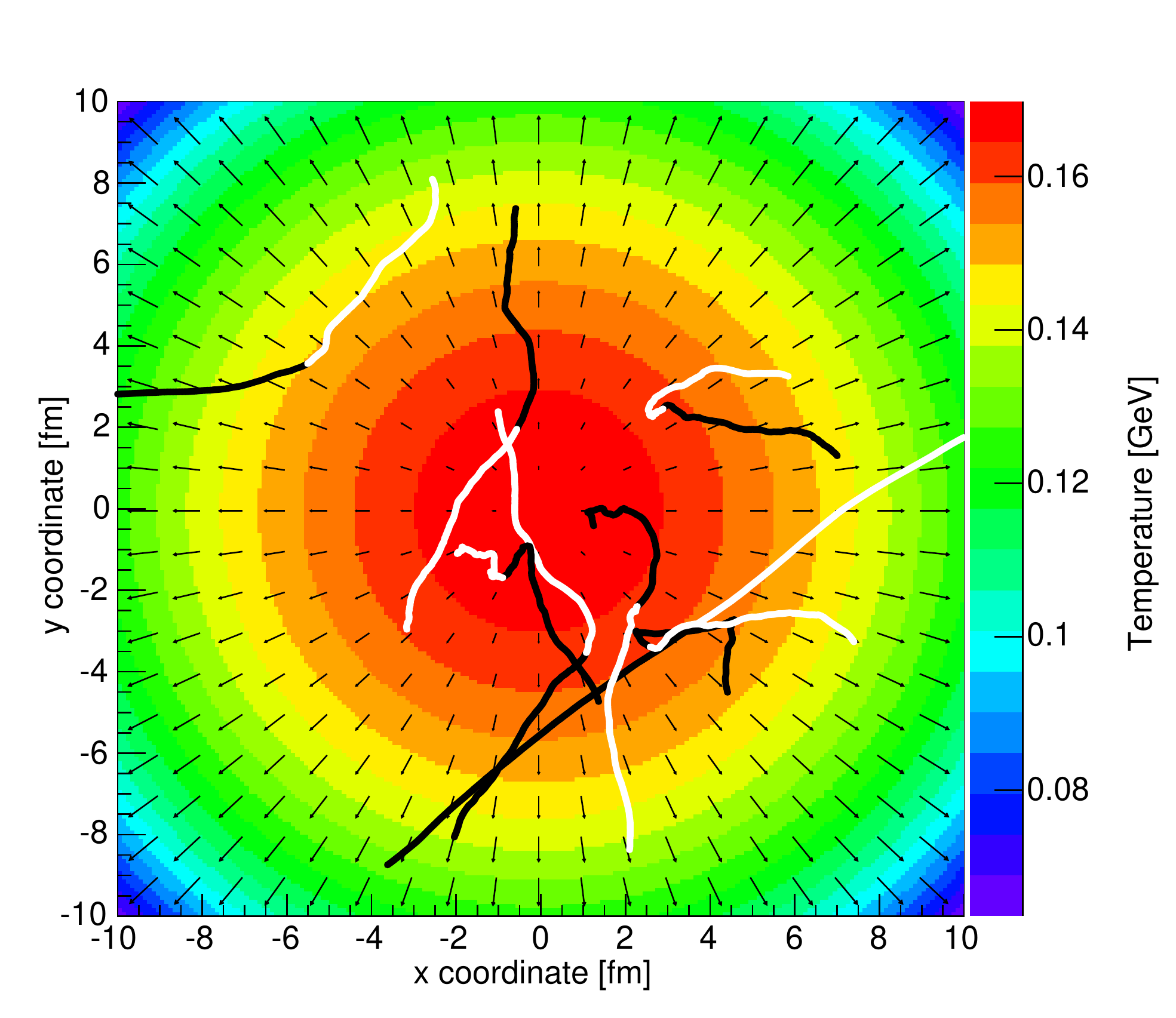}
  \caption{\label{fig:snapshot} Illustration of few $c\bar{c}$ pair
    trajectories in the expanding medium after 10 fm/$c$, with a
    diffusion parameter value of $D = 3/2\pi T$.}
\end{figure}
Figure~\ref{fig:snapshot} shows a visual record of the path traversed
by a few typical charm-anticharm pairs. The nuclear modification of
the $c$ and $b$ quark $\pt$ distributions is plotted in
Figure~\ref{fig:timesteps} for different ``snapshots'' during the
evolution of the system.
\begin{figure}[htb]
  \includegraphics[width=0.49\linewidth]{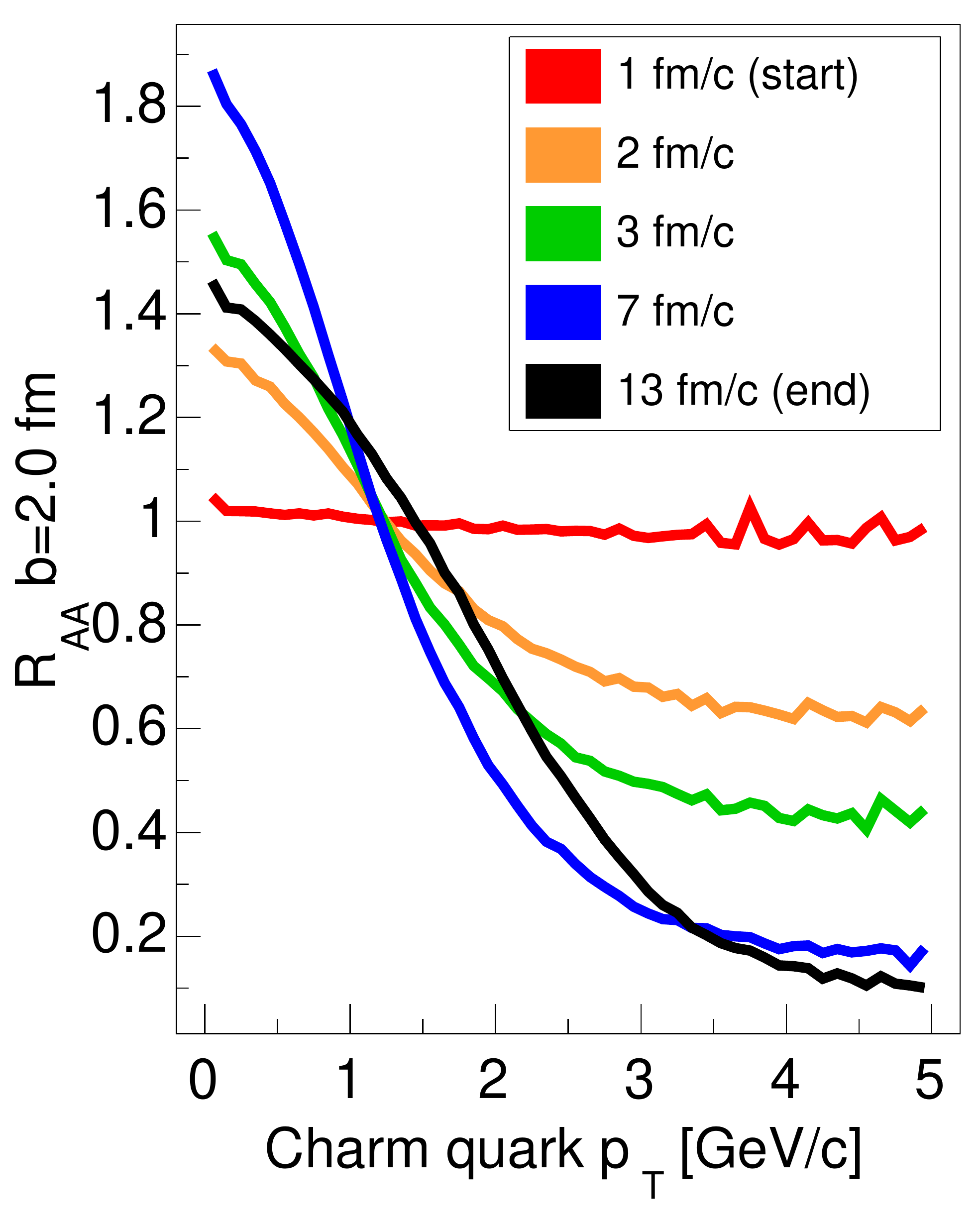} 
  \includegraphics[width=0.49\linewidth]{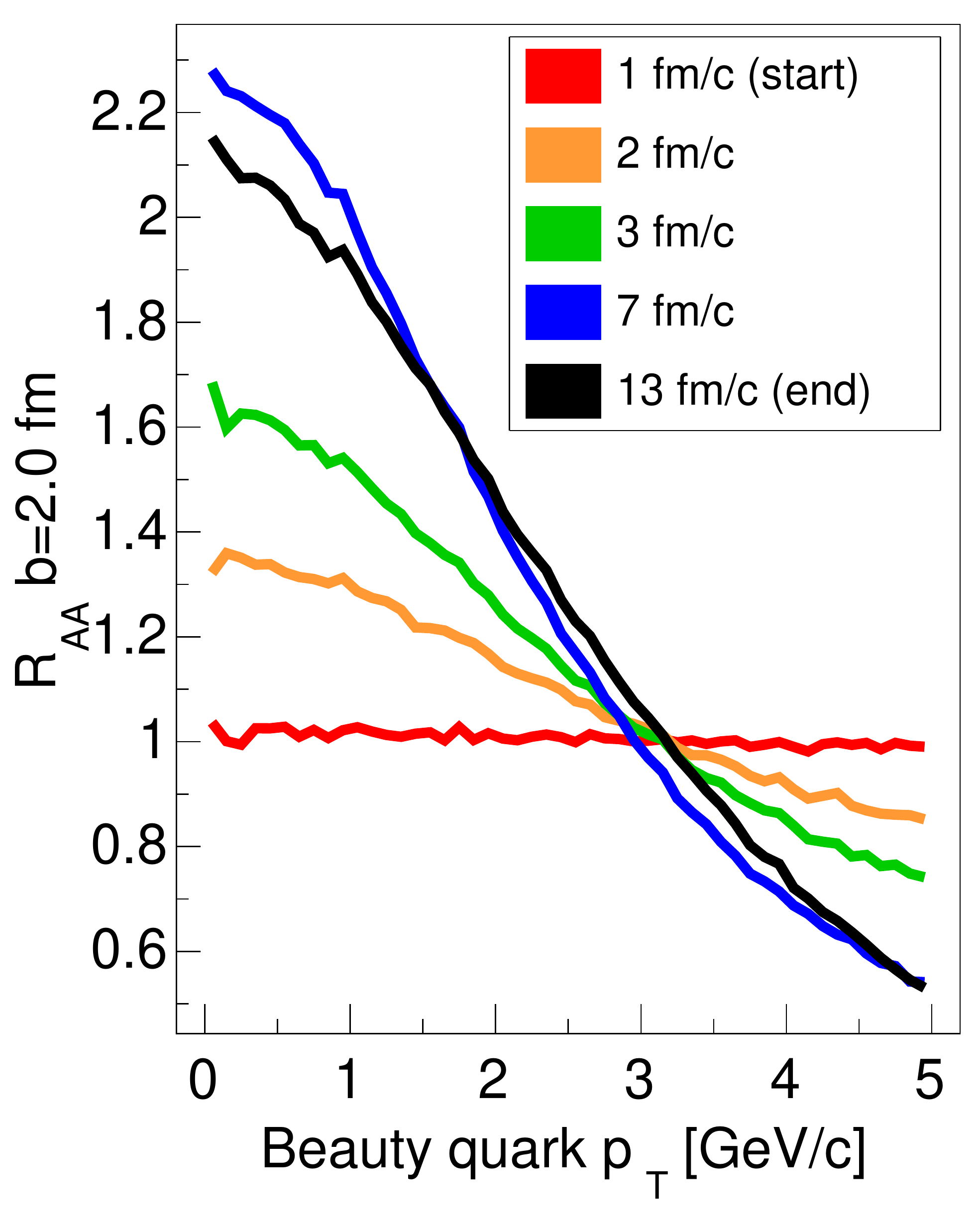} 
  \caption{\label{fig:timesteps} Charm quark (left) and beauty quark
    $\raa$ (right) at various times during the hydrodynamic evolution.}
\end{figure}
At the starting time of 1 fm/c, the $\raa$ is statistically consistent
with one. During the first 7 fm/c, the $\raa$ rises at low $\pt$, and
drops at higher $p_{T}$. After 7 fm/c, the trend reverses: $\raa$
moves downward at low $p_{T}$ and increases at intermediate $p_{T}$,
due to the strong radial flow velocities that have developed in the
medium. Since the initial heavy-quark $\pt$ distribution is much
harder than the thermal distribution of the quark-gluon plasma, the
Langevin drag term dominates over the diffusion term, pulling the $c$
and $b$ quarks to lower $p_{T}$. Given enough time in a static medium,
the heavy quarks would eventually follow a thermal distribution with
the medium temperature, as studied in~\cite{PhysRevC.84.064902}. It is
notable, however, that the final $\raa$ remains above $1.0$ at low
transverse momentum, in contrast to the suppression seen in the $D$
meson data and the blast wave result (Figure~\ref{fig:blast}).

The initial hydrodynamic results were produced assuming a constant
value of $\eta/s = 1/4\pi$ translated to the diffusion parameter using
the relation
\begin{equation}\label{eq:dpar}
D(T) = \frac{\eta}{s}\frac{6}{T}
\end{equation}
which is based on~\cite{Adare:2006nq}. For full consistency, the
hydrodynamic simulation should be modeled with a shear viscosity
following the same relationship as that applied to the heavy quarks,
but in order to isolate the effects of quark-medium interactions, the
hydrodynamical model always uses a constant shear viscosity such that
$\eta/s = 1/4\pi$ in all studies presented here.

\begin{figure}[htb]
 \includegraphics[width=0.49\linewidth]{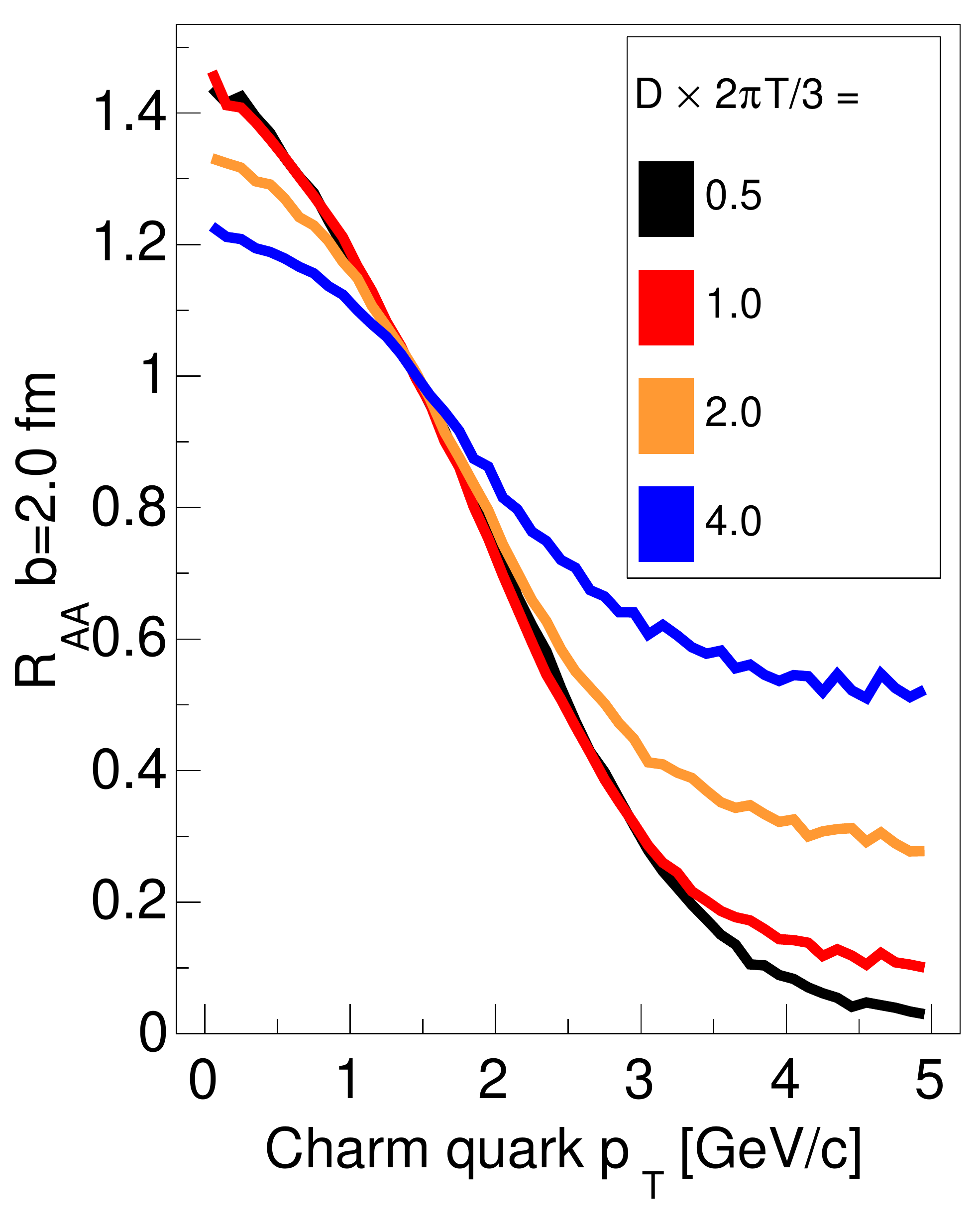} 
 \includegraphics[width=0.49\linewidth]{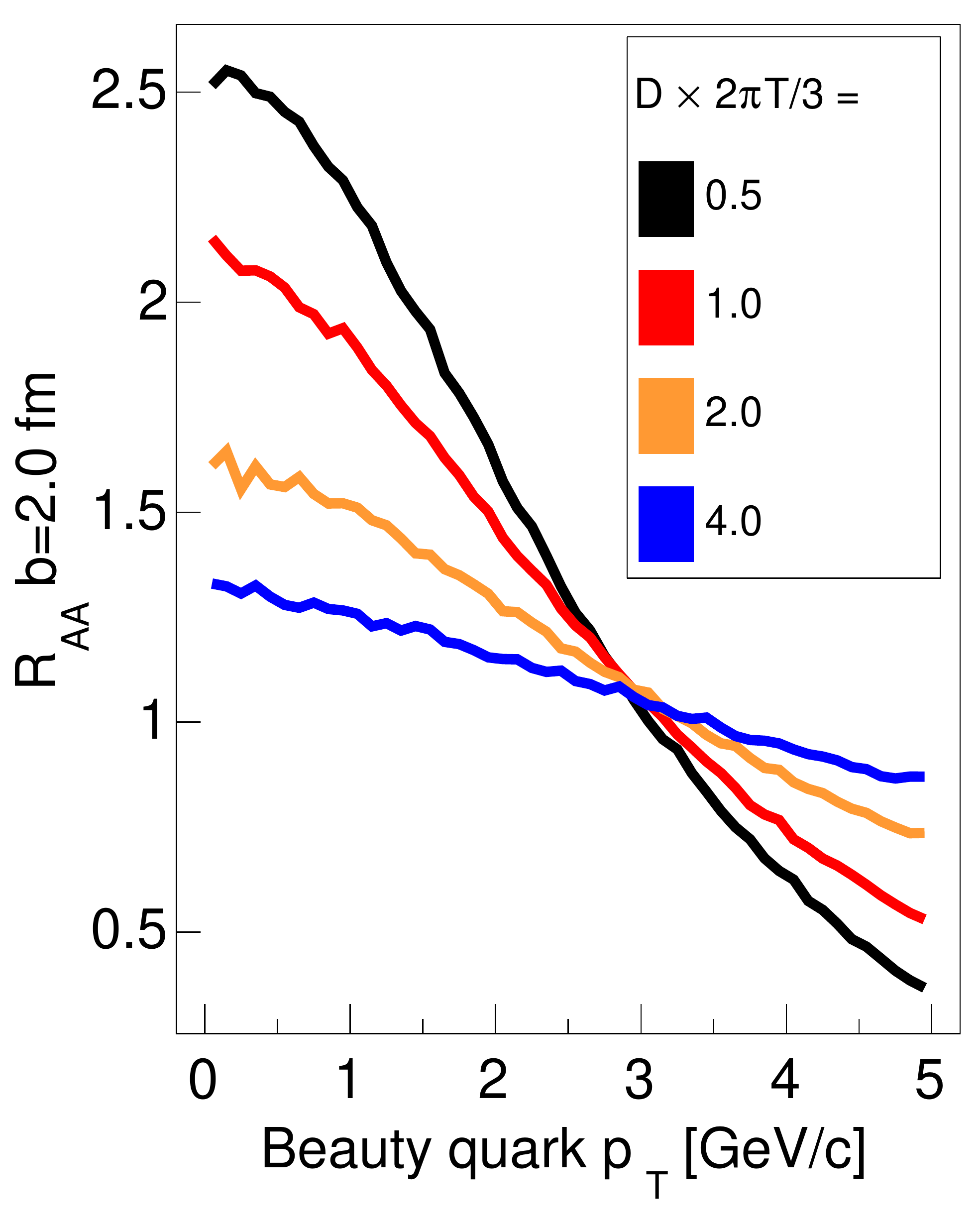} 
 \caption{\label{fig:scaleetas} Nuclear modification factor $\raa$ for
   charm quarks (left) and beauty quarks (right) at several values of
   the diffusion parameter D.}
\end{figure}

We explore the dependence of $\raa$ on diffusion strength by running
the calculation with a range of diffusion parameters $D$. We use the
correspondence of Eq.~\ref{eq:dpar} for $\eta/s$ equaling various
factors of $1/4\pi$, specifically $D = \{0.5, 1, 2, 4\} \times 3/(2\pi
T)$. The final $\raa$ curve for each $D$ value is shown in
Figure~\ref{fig:scaleetas} for $c$ quarks (left) and $b$ quarks
(right).
For charm quarks, the differences in $\raa$ for transverse momentum
below 2.0 GeV/c are quite modest ($\pm$ 10\%) despite an eight-fold
variation of the diffusion parameter. Even if the diffusion is made to
be extremely small by using, e.g. $D\times 2\pi T / 3 = 0.01$ (not
shown), $\raa$ remains above 1.0 at low $\pt$ for both species.

The low-$\pt$ charm quark $\raa$ is insensitive to the diffusion
strength because the initial drag and the late-stage radial push tend
to cancel one other. For beauty quarks, however, the low-$\pt$
enhancement is dramatically increased as $D$ is reduced, due to the
downward redistribution in $\pt$.  For the $b$ quarks, the late-stage
push is a weaker effect, leading to less cancellation against the
early-stage energy loss.

The balancing of early and late-time effects and the lack of ability
to achieve $\raa < 1$ at low $p_{T}$ led us to explore the temperature
dependence of the diffusion parameter. If $D$ increases at higher
temperatures (e.g. $\eta/s$ rises as a function of temperature above
the quark-gluon plasma transition temperature), then the early-time
drag will be weaker and the later time flow boost could result in a
depletion at low $p_{T}$.

\begin{figure}[htb]
  \includegraphics[width=0.9\linewidth]{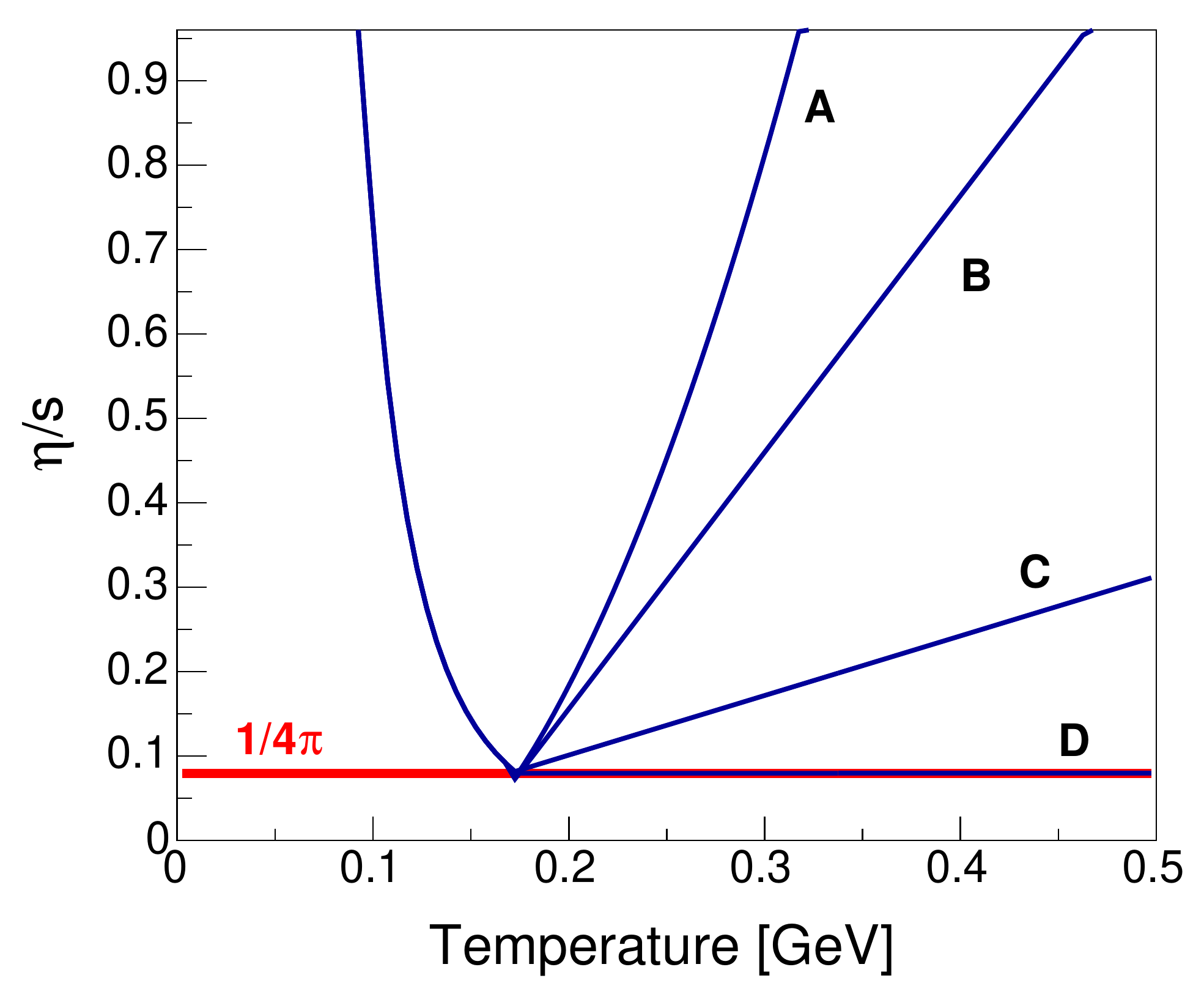} 
   \caption{\label{fig:abcd} Temperature dependence of
     $\eta/s$ for a variety of scenarios.}
\end{figure}

\begin{figure}[htb]
  \includegraphics[width=0.49\linewidth]{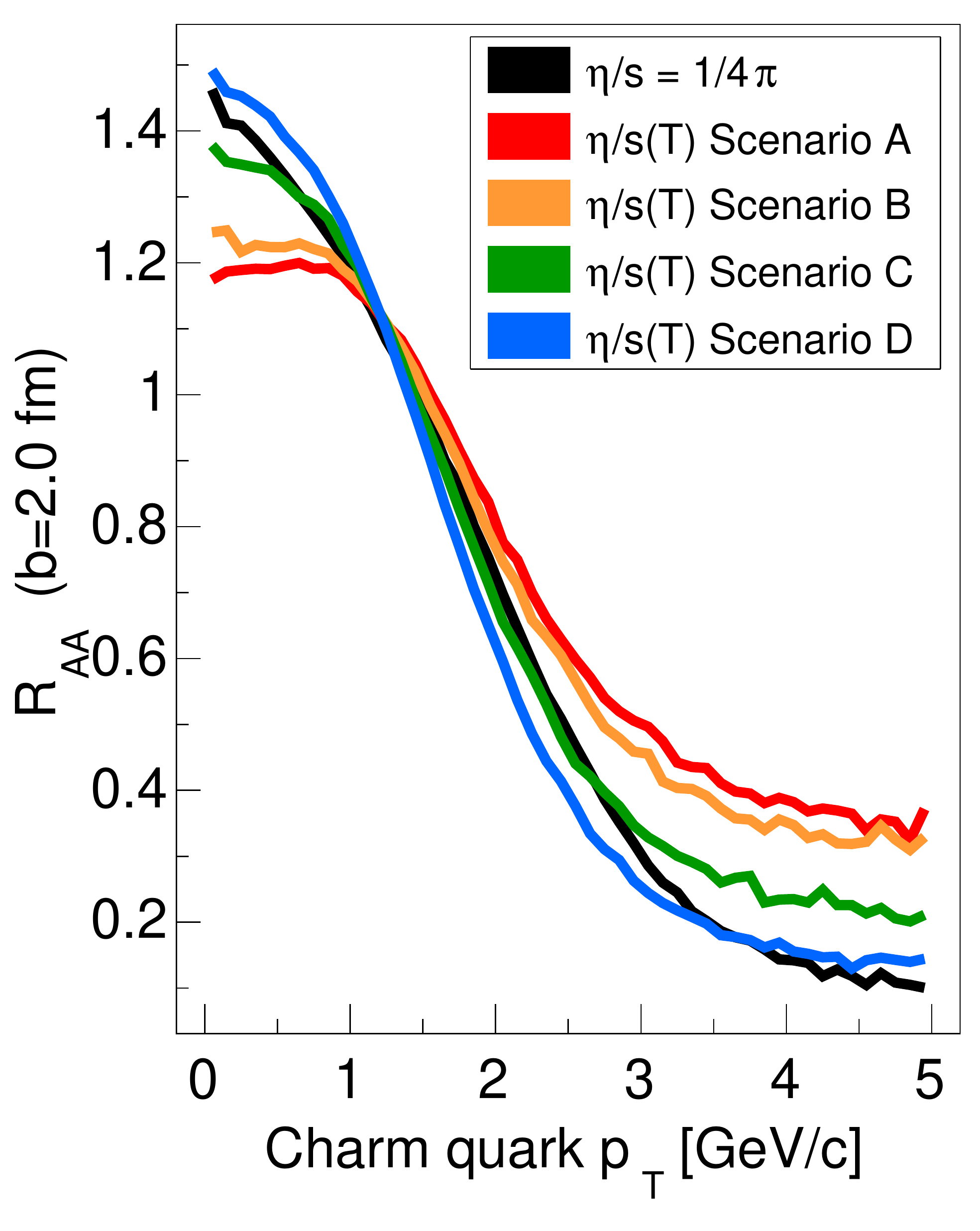}
  \includegraphics[width=0.49\linewidth]{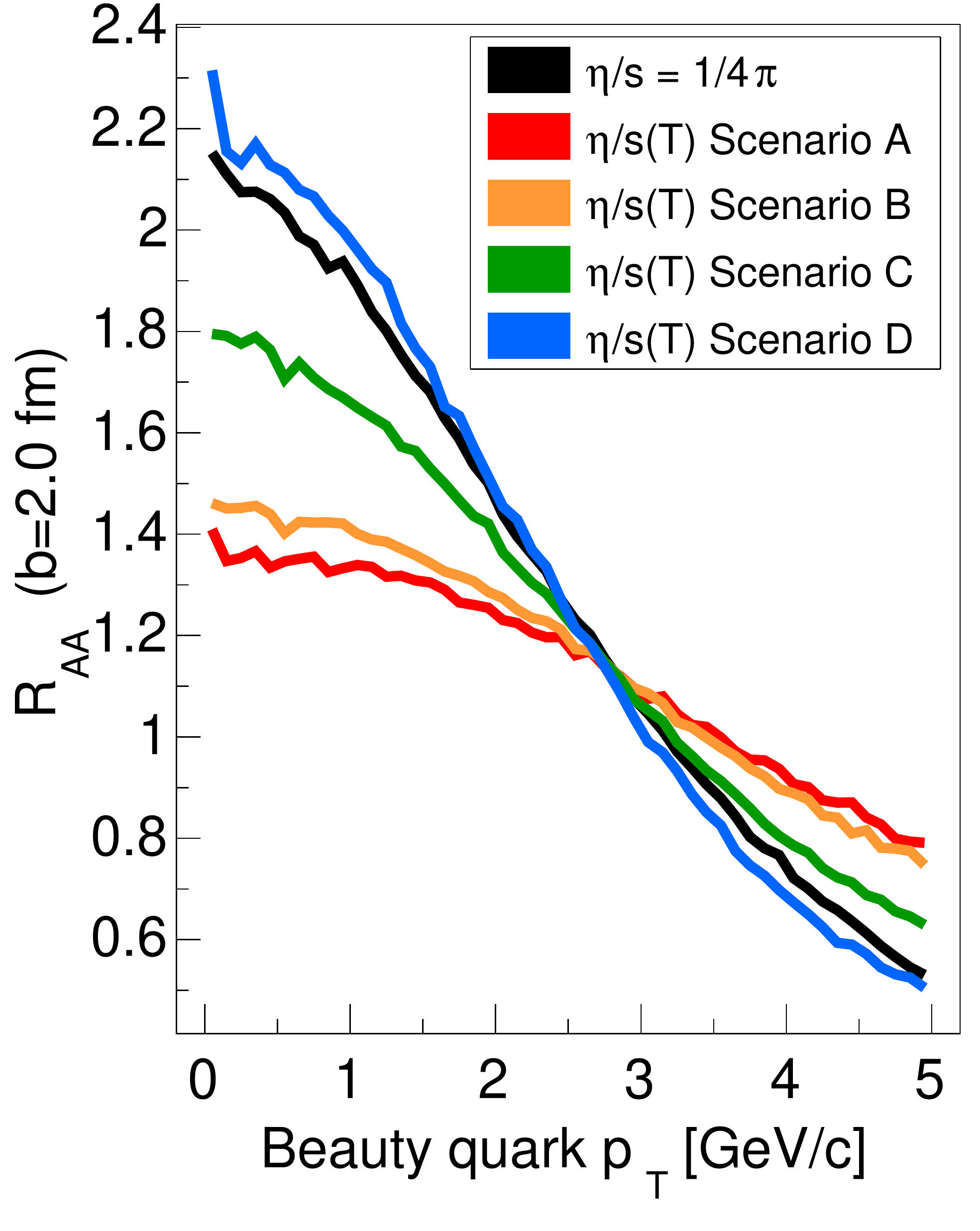}
  \caption{\label{fig:abcd_raa} Nuclear modification factors for charm
    quarks (left) and beauty quarks (right) for the set of $\eta/s
    (T)$ functions shown in Figure~\ref{fig:abcd}.}
\end{figure}

We have considered four different $\eta/s$ temperature dependence
scenarios from Ref.~\cite{Aidala:2012nz} (Figure~\ref{fig:abcd}),
which are converted to the diffusion parameter $D(T)$ using
Eq.~\ref{eq:dpar} for input to the Langevin calculation. Scenarios $B$
and $C$ are motivated by recent bulk hydrodynamic fits to the data at
RHIC and the LHC~\cite{PhysRevC.83.054910,PhysRevLett.106.212302,
  PhysRevLett.110.012302,Nagle:2011uz}.  We note that the temperature
dependence of the diffusion parameter from Ref.~\cite{Moore:2004tg}
was calculated perturbatively, and here we just phenomenologically
parameterize the lower temperature ($T < 500$ MeV) dependence. The
results from the four scenarios for the $c$ and $b$ quark nuclear
modification factor are shown in Figure~\ref{fig:abcd_raa}.

It is notable that none of the scenarios $A$-$D$ lead to a depletion
of charm quarks at low transverse momentum, despite the large
variation of diffusion strength at high-temperature. This suggests that
the strong $\raa$ ``hump'' in Figure~\ref{fig:blast} is not likely to
be from quark-medium interactions alone. In the blast-wave model, the
heavy quarks are distributed over the entire transverse plane such
that a large fraction are positioned at large radii, where the
late-stage hydrodynamic push is largest.

\begin{figure}[htb]\centering
  \includegraphics[width=0.9\linewidth]{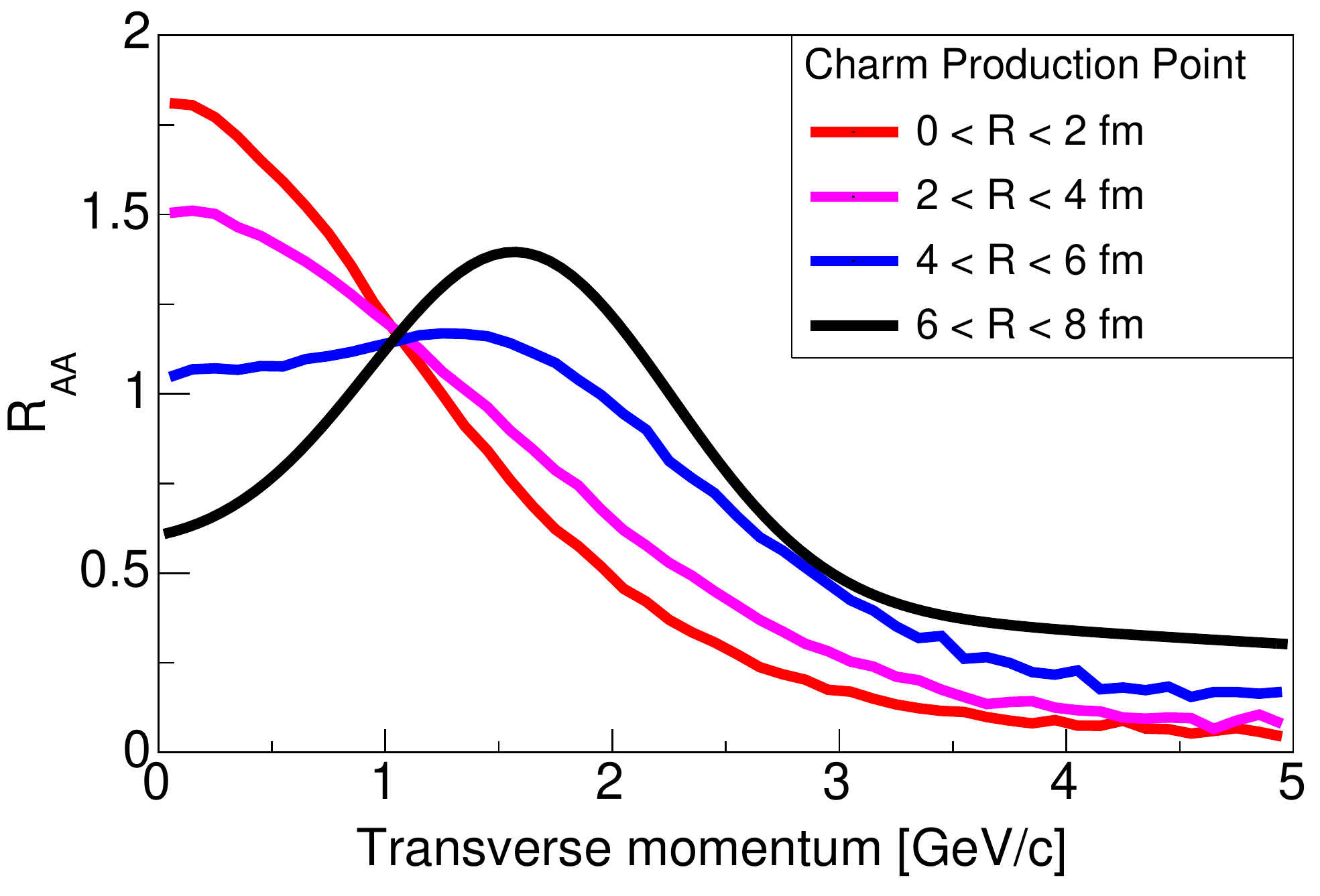}
  \caption{\label{fig:rslices} $\raa$ for charm quarks produced within
    four different radial intervals using $D = 3/2\pi T$.}
\end{figure}

To demonstrate the dependence of nuclear modification on initial quark
radial positions $R$, Figure~\ref{fig:rslices} shows $\raa$ for charm
quarks originating in several different $R$ selections.
Only when all charm quarks originate at $R > 6$ fm does the Langevin
charm $\raa$ qualitatively reproduce the shape found in
Figure~\ref{fig:blast}. Since most quarks originate within $4$ fm of
the medium centroid in any realistic central Au+Au model, the
initial-state geometric configuration appears unlikely to play a large
role in determining the shape of $\raa$. It has been suggested that
pre-equilibrium radial flow may redistribute the heavy quarks outward
and impart a significant radial velocity to the heavy quarks.  We did
study the effect of pre-equilibrium flow from
Ref.~\cite{vanderschee2013} in our Langevin calculation, but did not
observe any qualitative change to the results presented here.

We note that in Figure 4 of Ref.~\cite{Moore:2004tg} for the smallest
diffusion parameter considered, the charm $\raa$ does turn down at low
$\pt$, though never decreasing below one. Running our calculation also
for $b = 6.5$ fm (midcentral) Au+Au and with identical parameters, we
qualitatively reproduce these results. Despite smaller fluid
velocities in more peripheral events, it is more likely for the charm
quarks to be located near the surface of the medium.

To recapitulate, we have found that in central Au+Au events, no
moderate value for the Langevin diffusion parameter, nor any realistic
distribution of heavy quark initial positions, nor pre-equilibrium
flow, is capable of producing the low-$\pt$ heavy-quark $\raa$ values
such as those observed for $D$ mesons in Figure~\ref{fig:blast}. It is
possible that the low-$\pt$ heavy-flavor meson $\raa$ is not primarily
due to physics occurring at the partonic stage, but rather hadronic
mechanisms such as coalescence~\cite{PhysRevLett.110.112301}.

We have seen that $\raa$ for charm quarks with $\pt<$ 2 GeV/c does not
reflect a strong dependence on the diffusion coefficient in the
high-temperature regime. A quantity that is potentially more sensitive
to early-time dynamics is the distribution in relative azimuth $\dphi$
for heavy quark pairs, which has been studied previously in
Refs.~\cite{Molnar2006,Zhu:2007ne,Nahrgang:2013saa,Tsiledakis:2009qh}.

\begin{figure}[htb]
  \includegraphics[width=0.49\linewidth]{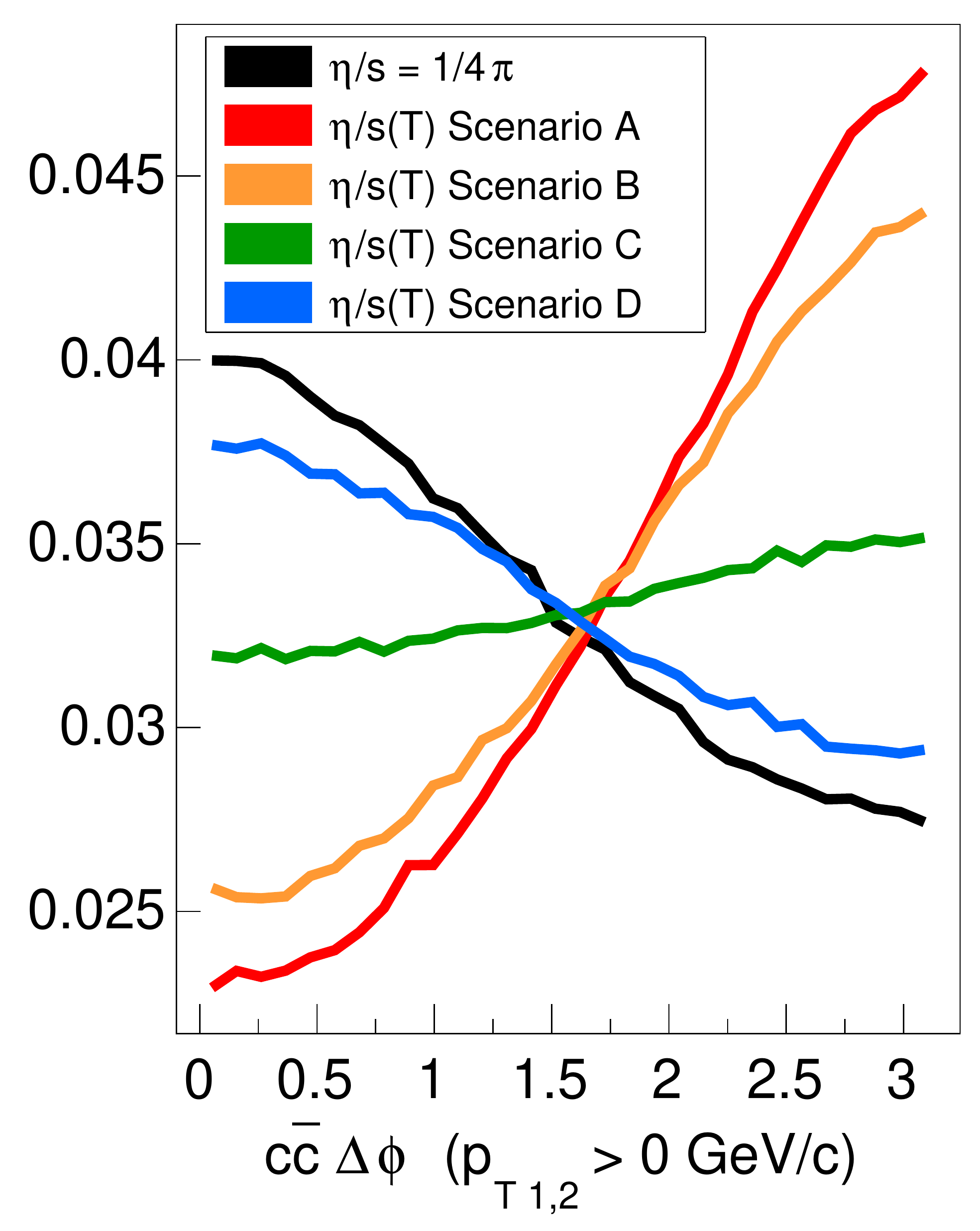}
  \includegraphics[width=0.49\linewidth]{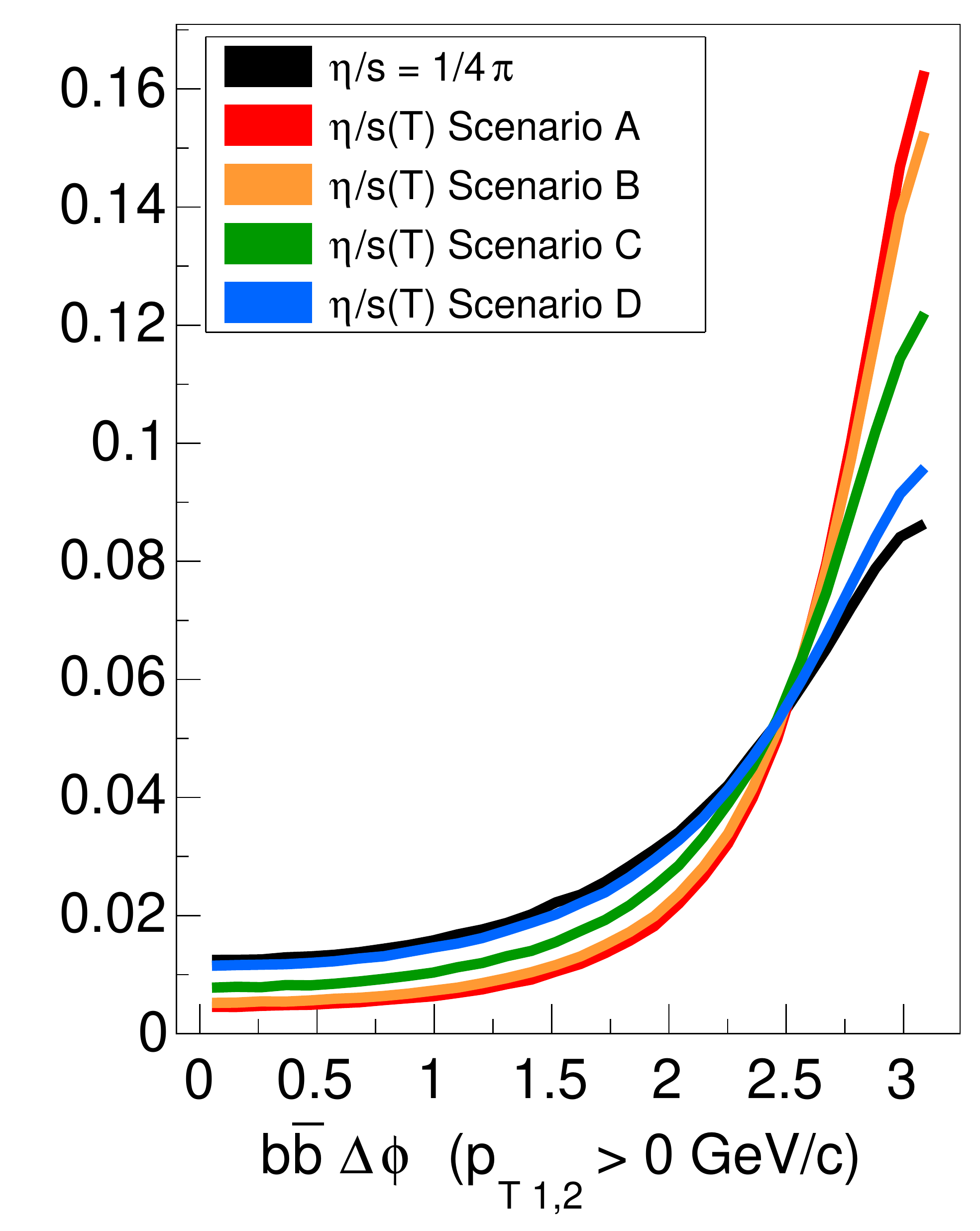}\\
  \includegraphics[width=0.49\linewidth]{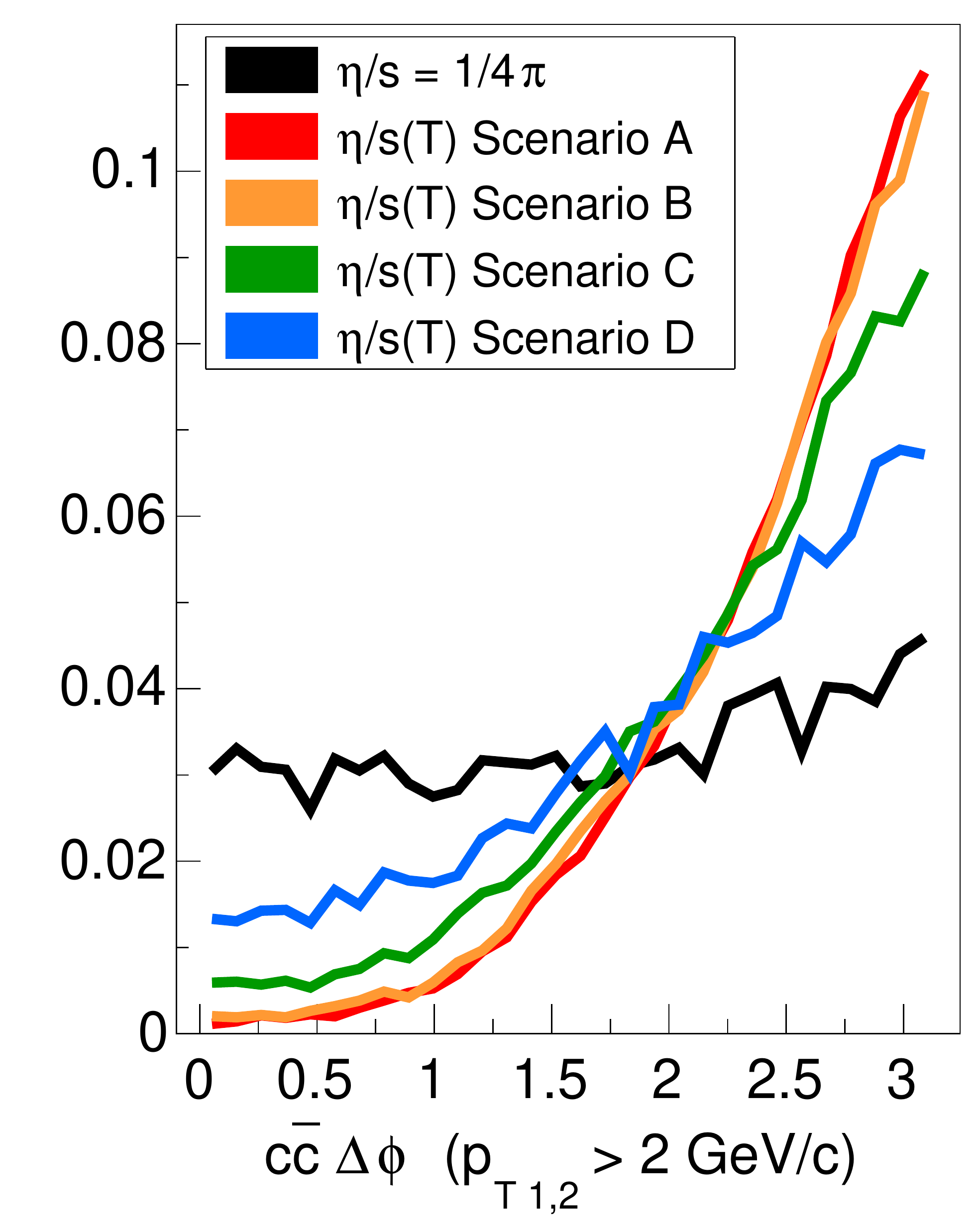}
  \includegraphics[width=0.49\linewidth]{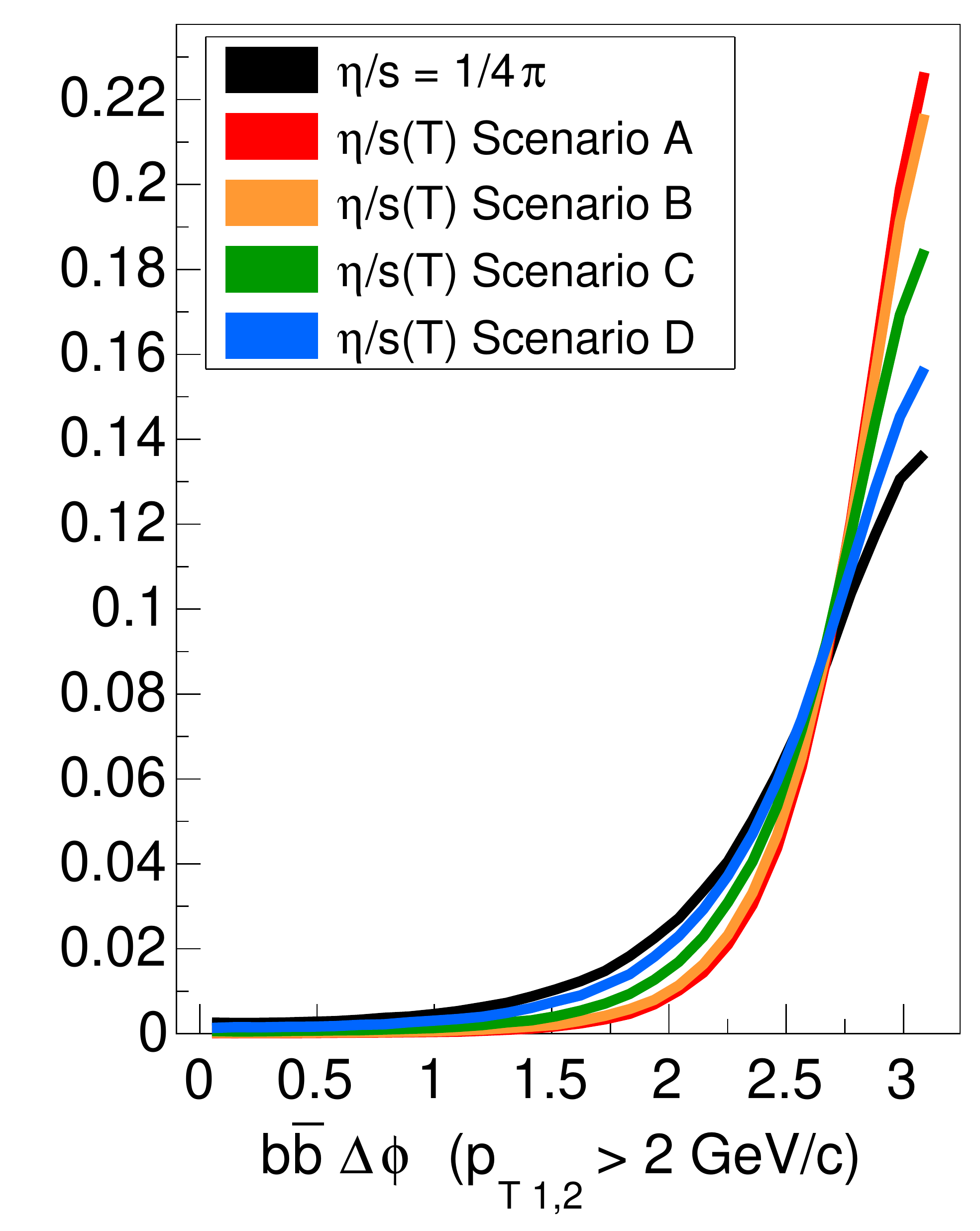}
  \caption{\label{fig:abcd_dphi} $\dphi$ distributions for charm
    quarks (left) and beauty quarks (right) for the set of $\eta/s
    (T)$ functions shown in Figure~\ref{fig:abcd}. The top row
    includes $\qqbar$ pairs at all momenta, and the bottom row
    includes $\qqbar$ pairs where both quarks have $\pt > 2$ GeV/c.}
\end{figure}

In striking contrast to $\raa$, the $\ccbar$ $\dphi$ distributions
shown in Figure~\ref{fig:abcd_dphi} reflect a very strong sensitivity
to variations in high-temperature diffusion.  The beauty quarks have
the opposite sensitivity, where the $\raa$ is more sensitive to the
early time stage and the $\dphi$ is relatively insensitive.

When $D$ is large at high temperature, the $\ccbar$ angular
correlation is peaked at $\dphi = \pi$. This feature is expected for
weak early-time quark-medium interactions, where the initial
back-to-back kinematics are preserved.  When $D$ is small, the
$\ccbar$ angular distribution exhibits a distinct near-side
correlation. This is due to (a) strong initial scattering and drag
that slows the quarks and destroys their initial opposing
trajectories, and (b) the late-stage radial push that acts to
collimate the quark-antiquark pairs -- as seen for example in the top
right quadrant in Figure~\ref{fig:snapshot}. For the $\bbbar$ pairs,
however, the initial energy loss is considerably smaller than for
charm quarks at comparable momenta, as shown in
Figure~\ref{fig:scaleetas}, thus retaining the away-side dominated
azimuthal pair distribution. 

In summary, the Monte Carlo Langevin framework, coupled with a
time-dependent viscous hydrodynamic medium model, provides a useful
tool for studying the space-time evolution of interactions between
heavy quarks and the thermal medium. Stochastic scattering and viscous
drag lead to high-$\pt$ suppression, as well as an enhancement of
particles at intermediate $\pt$.  However, late-stage hydrodynamic
expansion is insufficient to cause $\raa < 1$ for very low $\pt$ heavy
quarks when a realistic initial geometry is used. Hadronization
mechanisms, such as coalescence, may be relevant in explaining
the low-$\pt$ suppression observed in heavy-flavor mesons.  These
calculations indicate that azimuthal correlations involving $\ccbar$
pairs are more sensitive to the diffusion strength than $\raa$.  In
contrast, the heavier $\bbbar$ has greater sensitivity via the $\raa$
than via correlations.  Next steps include identifying the specific
optimal experimental observables reflecting the underlying heavy quark
final distributions.
 


\begin{acknowledgments}
  We gratefully acknowledge useful discussions with Joerg Aichelin,
  Steffen Bass, Shanshan Cao, Matt Luzum, and Krishna Rajagopal.  AMA
  and JLN acknowledge support from the United States Department of
  Energy Division of Nuclear Physics grant DE-FG02-00ER41152. PR
  acknowledges support from DOE award No. DE-SC0008027 and Sloan Award
  No. BR2012-038. MPM acknowledges support from the Los Alamos
  National Laboratory LDRD project 20120775PRD4.
\end{acknowledgments}



\bibliography{langevin}

%


\end{document}